\documentclass[twocolumn,epjc3]{svjour3}          
\usepackage{newtxtext,newtxmath}
\usepackage{physics}
\usepackage{lineno}

\RequirePackage[T1]{fontenc}

\smartqed  

\RequirePackage{graphicx}
\RequirePackage{flushend}
\RequirePackage[numbers,sort&compress]{natbib}
\RequirePackage[colorlinks,citecolor=blue,urlcolor=blue,linkcolor=blue]{hyperref}

\journalname{Eur. Phys. J. C}

\begin{document}

\title{Imaging the radio-wave emission from extensive air showers}

\author{Juan Ammerman-Yebra\thanksref{e1,addr1,addr2}
        \and
        Harm Schoorlemmer\thanksref{addr1,addr2} 
        \and
        Anne Timmermans\thanksref{addr3,addr4}
        \and
        Sebastian Achim Mueller\thanksref{addr4}
}

\thankstext{e1}{e-mail: juan.ammermanyebra@ru.nl}

\institute{IMAPP, Radboud University Nijmegen, Heyendaalseweg 135, 6525 AJ Nijmegen, The Netherlands\label{addr1}
          \and
          Nationaal Instituut voor Kernfysica en Hoge Energie Fysica (NIKHEF), Science Park, Amsterdam, The Netherlands\label{addr2}
          \and
          Fellow of the International Max Planck Research School for Astronomy and Cosmic Physics at the University of Heidelberg (IMPRS-HD)\label{addr3}
          \and
          Max-Planck-Institute for Nuclear Physics, Heidelberg, Germany\label{addr4}
}

\date{Received: date / Accepted: date}

\maketitle

\begin{abstract}
We propose a new way to observe cosmic-ray-induced air showers by imaging the radio emission. With simulations we demonstrate key features for imaging the radio-wave emission from air showers, which show similarities to the well-established atmospheric imaging Cherenkov technique in gamma-ray astronomy. In addition, we find that imaging the emission with a camera, consisting of multiple antennas, resolves emission that is not accessible to a single antenna. Pursuing this technique, with a camera operating in the GHz frequency domain, might be beneficial ultra-high-energy gamma-ray astronomy and other studies that include detailed  observations of air showers.
\end{abstract}

\maketitle
\section{Introduction}

Energetic cosmic particles initiate particle cascades in the Earth’s atmosphere, known as extensive air showers (EASs). The propagation of EASs through the atmosphere causes several emission processes in the electromagnetic spectrum: At visible and ultraviolet wavelengths, Cherenkov and fluorescence light are emitted, while at radio wavelengths the varying charge distributions in the EAS cause emission. On the detection side, the way these emissions are observed can be classified into two categories: imaging, in which the emission of the air shower is recorded by a pixelated camera, and non-imaging, in which all the integrated emission of the air shower is recorded by a single sensor. Examples of imaging detectors are Fluorescence Telescopes (used at the Pierre Auger Observatory \cite{AugerFD} and at the Telescope Array\cite{TAFD}) and Imaging Atmospheric Cherenkov Telescopes (like H.E.S.S. \cite{HESS},  \cite{MAGIC} and VERITAS \cite{VERITAS}). In the case of non-imaging observations, we have radio-wave detector arrays, that typically operate in the 30-80\,MHz frequency band (for example at the Pierre Auger Observatory \cite{AUGER_RD,AERA} and at the LOw Frequency ARray  LOFAR \cite{LOFAR}), and non-imaging Cherenkov detector arrays (for example HiSCORE \cite{Budnev:2025wuc}).\\
In this work, we investigate the features of imaging observations of radio-wave emission from air showers. There have been several prototype telescopes (AMBER \cite{AMBER}, CROME \cite{CROME}, and MIDAS \cite{MIDAS}) with imaging cameras with the main objective of detecting isotropically emitted molecular brems\-strah\-lung, which would have led to the radio equivalent of fluorescence telescopes. However, from the results by CRO\-ME \cite{CROME}, it was clear there is radio-wave emission in the GHz frequency domain that is emitted in a small cone around the axis of the air shower, similar to the emission detected at a lower frequency band by radio-wave detector arrays. Here, we investigate the features of imaging this radio-wave emission, which would correspond to the radio-wave equivalent of the IACTs. The focus of this article is on the implications of the observed emission, while the details of implementing a realistic system will be discussed in another article. 
 
\section{Imaging the emission from air showers}
To study the imaging of radio-wave emission from air showers, we modified the ZHAireS code \cite{AlvarezMuniz2012}. ZHAireS calculates from each charged particle track in the air shower the contribution to radio-wave emission through the ZHS algorithm \cite{zas1992electromagnetic, Alvarez-Muniz:2010wjm}. In its normal mode, it sums the contributions from each particle track to obtain a total electric field at an antenna location. We modified ZHAireS such that at one location, you can have several antennas, each with a limited field-of-view, like you would have for a pixel in a camera. In each \textit{pixel} we now sum only the contributions from the pixel’s field-of-view. Since we study the general features of imaging, we use an idealized camera, in which the field-of-view of pixels does not overlap. In the following, we will term a \textit{camera} for the electric field sensor that performs pixelized observation and an \textit{antenna} when referring to a single electric field sensor at a location. We define a viewing angle $\theta$ as the angular distance with respect to the viewing direction of the center of a camera. Implementing imaging in a practical device would include the implementation of a reflecting surface to focus the emission on the camera. However, to  decouple the study of the potential of the technique from practical design implementations, we are currently not considering the impact of a reflecting dish. 

Before going to the case of a two-dimensional grid of pixels, we highlight some features by pi\-xe\-li\-zing only one dimension. We simulate a vertical air shower and pixel the field of view only in the zenith angle, i.e. each pixel accepts contributions from a specified zenith angle range. The simulated air shower is initiated by a 10\,PeV proton that propagates through the atmosphere with a 0.4\,G horizontal magnetic field aligned with the $x$ direction.

\begin{figure}[!ht]
\includegraphics{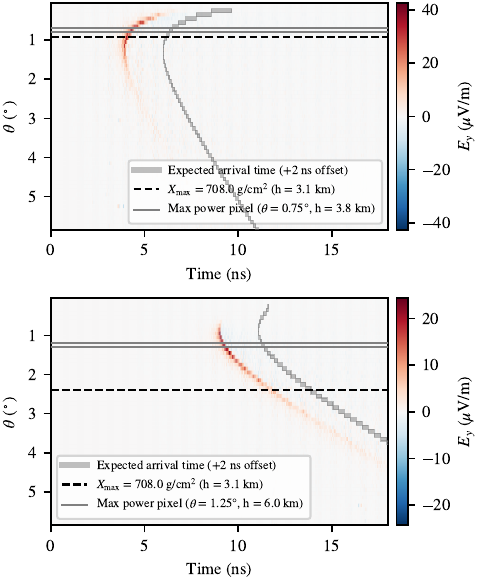}
\caption{\label{fig:arrival_time} Signal strength (color axis) and arrival time of the radio emission (with an offset of 2\,ns for better visualization) while pixelized in the zenith angle for two cameras, one placed at ($x=50\,\text{m},y=0\,\text{m})$ (top) and one at ($x=130\,\text{m},y=0\,\text{m})$ (bottom). The black dashed line indicates the location of the shower maximum and the pixel with highest integrated power is highlighted by two gray horizontal lines. 
}  
\end{figure}
Figure~\ref{fig:arrival_time} shows the resulting electric field for two cameras at different locations, which clearly illustrate that with a camera you can image which part of the longitudinal air shower development is responsible for the emission. This is in contrast to the traditional approach, where a single antenna at a location measures the emission integrated over the full longitudinal development. The properties of the air shower development are subsequently obtained indirectly through comparisons of the emission recorded at different antenna locations.
The time structure as a function of the viewing angle of the main pulse in each pixel is fully understood by calculating the propagation time from the shower axis to the camera location. Imaging the air shower at different distances from the shower axis alters the time profile in a predictable way, which means that in principle, by fitting the emission propagation time model to the observations, the distance of the camera location to the air shower axis can be obtained.\\
We now continue by generating a camera, which has a form factor similar to the cameras used in IACTs. The diameter of the total field of view of the camera is 6$^\circ$ and we populate the focal plane with 823 pixels. In simulations, there is no constraint on the frequency range for each pixel; however, to restrict ourselves to pixel sizes that result in a camera that is feasible to construct, we will evaluate the power of the emission in the frequency band between 2 and 4\,GHz. In addition, we will evaluate the power of the emission corresponding to the  polarization expectation of the geomagnetic emission.\\
\begin{figure*}[!ht]
    \centering
    \includegraphics[width=\textwidth]{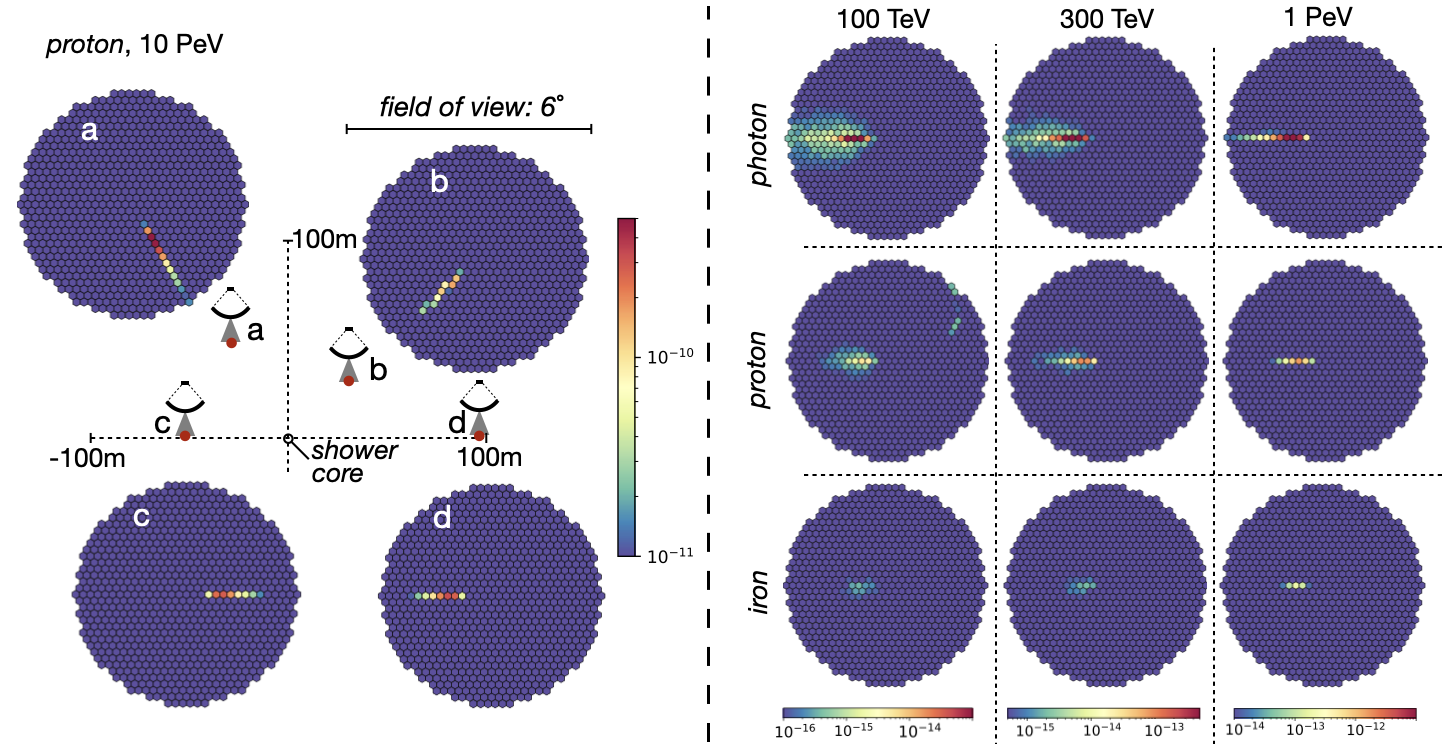}    
    \caption{Left: Imaging the power of the radio-wave emission of a simulated 10\,PeV proton-induced air shower between 2 and 4\,GHz (colorbar units in (ns V$^2$/m$^2$)  ). The cameras are placed at different locations in a horizontal plane and point straight up. The proton comes in from zenith. Right: Images of air showers induced by different type of cosmic  primaries as a function of their energy.}
    \label{fig:camera_geometry}
\end{figure*}
In Figure \ref{fig:camera_geometry} (Left), we show the camera view of a simulated air shower, initiated by a vertical 10\,PeV proton, for four different camera positions to illustrate geometry effects on imaging an air shower from different locations. It is clear that the geometry follows the same imaging features as for IACTs, with a one-to-one correspondence of the image major axis aligning with the azimuthal location of the camera in the horizontal plane. We also see that the image is roughly aligned with a single line of pixels, indicating that the radio-wave emission is emitted close to the axis. Together with the possibility of using the timing to constrain the distance to the shower axis (see Figure \ref{fig:arrival_time}), this will potentially allow for accurate reconstruction of the air shower geometry from a single camera observation.\\
In Figure \ref{fig:camera_geometry} (Right) we show the impact on the images depending on particle type and their energy. We observe that the emission for the photon showers is significantly more intense and focused along the shower axis than hadron showers with the same primary particle energy.  The more nucleons the primary has, the less coherent emission is observed from the air shower.  This is most likely caused by the higher multiplicity of sub showers, which results in a spread of electromagnetic cascades that reduce the charge density. Similar, for all primaries, both the total number of charges and the relative charge density are reduced with decreasing energy of the primary. A photon primary will cause a single electromagnetic cascade, hence the image remains focused along the shower axis even at 100\,TeV. The observed reduction of intensity, and spread over pixels, goes together with a general loss of polarization. For photons-induced air showers the dominant contribution to the observed emission remains polarized as expected for the geomagnetic emission down to energies of about 30\,TeV.\\

\section{Observing additional emission by imaging}
Since radio-wave emission emitted at a certain location can interfere with emission coming from other locations and at different times before it is observed, we might expect a camera to observe emission contributions beyond those detectable by a single antenna. By applying the camera observations, we identified two additional contributions to the observed emissions that would be missed by using a single antenna.\\ 
\begin{figure}
    \centering
    \includegraphics[width=0.95\linewidth]{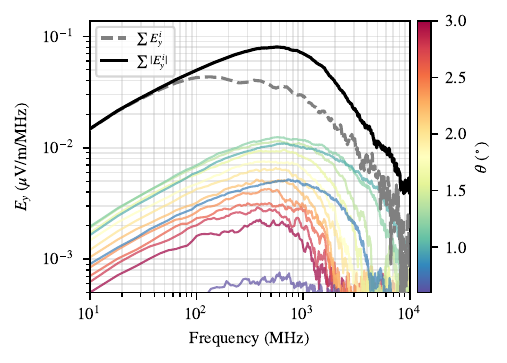}
    \caption{Amplitude spectra for pulses observed by individual pixels along the shower axis shown by the colored lines, the black solid line shows total amplitude spectra observed by a camera, while the gray dashed line shows the total amplitude spectra for a single antenna. Signals were obtained using camera ``\textit{d}'' displayed in Figure \ref{fig:camera_geometry} (left).}
    \label{fig:spectra}
\end{figure}
The most important effect is a boost of emission at higher frequencies. The coherence condition for a moving emitter is set by the time that the emitter spends within the field-of-view of the pixel, which leads to coherency at higher frequencies for smaller pixel sizes, this effect can be seen in Figure\,\ref{fig:spectra}. When summing the amplitude spectra of individual pixels, we get a total amplitude spectrum as observed by a camera. This is compared to the spectra of a single antenna, which is obtained by first summing in the time domain all the pulses and then calculating the amplitude spectrum. The camera total spectrum keeps rising up to $\sim$700\,MHz while the single antenna spectrum decreases just above 100\,MHz.\\
\begin{figure}
    \centering
    \includegraphics[width=0.95\linewidth]{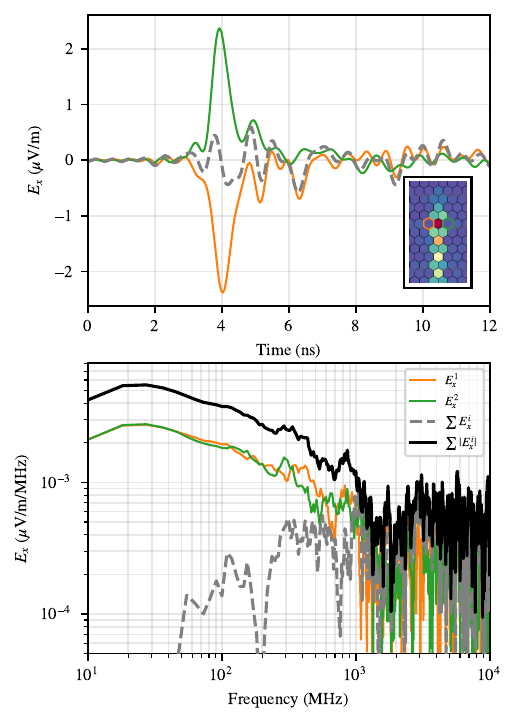}
    \caption{Destructive polarization interference of emission from opposite sides of the air shower axis. Time traces have been filtered between 0 and 2000 MHz to reduce the noise for better visualization.}
    \label{fig:new_emission}
\end{figure}
Another, more subtle contribution, is observed by evaluating the emission of two pixels on opposite sides of the shower axis as shown in Figure~\ref{fig:new_emission}. The camera is placed at x=0\,m and y=50\,m, such that Askaryan and the geomagnetic emission are both polarized in the $y$-direction\cite{geo_charge}. We find emission in $x$-polarization for the two pixels, which has an amplitude that is in the order of 5\% when comparing to the on-axis total emission. We observe that the polarity flips with the side of the shower axis. A single antenna observes the sum of these two contributions and they would interfere destructively. However, the emission can be resolved with a camera. We note that this is a previously unknown contribution to the radio-wave emission which is compatible with a velocity component of charged particles in the transverse plane.\\

\section{Conclusion}
Imaging the radio-wave emission from simulated air showers provides insights into both the emission processes and the prospect for experimental implementation. Compared to observing with a single antenna, we identified a boost of the power towards higher frequencies facilitating observations with receivers that observe in the GHz range. Within this frequency band, cameras have already been operated with \textit{phased array feeds} for astronomical surveys (like Apertif \cite{van2022apertif} and ASKAP\cite{ASKAP}). Like in radio astronomy and imaging Cherenkov telescopes, a camera can be combined with a large reflector focusing the emission onto a focal plane. The sky temperature above one GHz is reduced to a few Kelvin (see for example \cite{Fixsen_2011}), which is insignificant compared to the traditional observing band (30--80\,MHz) where the sky temperature is several thousands Kelvin (see for example \cite{45MHz}). This means that thermal background noise will be dominated by the system noise temperature, which is typically in the range of 70--90\,K\cite{CROME, van2022apertif}. In addition, using a radio telescope, the emission is observed in the plane of polarization, unlike for a free-standing antenna where the emission comes in from different angles depending on the air shower geometry. 
Since radio-waves are hardly absorbed nor scattered in the atmosphere, and observable day and night, an eightfold boost in exposure is reachable compared to the IACTs. In addition, the angular resolution that might be obtainable by imaging the radio emission from air showers might exceed that of imaging Cherenkov telescopes, as the emission is not scattered in the atmosphere, it maintains its spatial and time--structure. \\
For a realistic estimate of the energy threshold for detecting cosmic particles, and realistic estimates of other performance parameters, a full model of a radio telescope is needed. Such a study is outside the scope of this work, but will be presented in another paper. 
In addition, making proof-of-principle measurements are needed to confirm the features of the radio emission that are presented in this work.

\section*{Acknowledgments}
J. Ammerman-Yebra and H. Schoorlemmer are funded by the European Union. Views and opinions expressed are however those of the author(s) only and do not necessarily reflect those of the European Union or the European Research Council Executive Agency. Neither the European Union nor the granting authority can be held responsible for them. This work is supported by ERC grant (CR-INTERFEROMETRY 101170979). 

\bibliographystyle{spphys}
\bibliography{main}

\end{document}